\begin{filecontents*}{example.eps}
gsave
newpath
  20 20 moveto
  20 220 lineto
  220 220 lineto
  220 20 lineto
closepath
2 setlinewidth
gsave
  .4 setgray fill
grestore
stroke
grestore
\documentclass[onecollarge,natbib]{svjour2}
\bibpunct{[}{]}{;}{n}{}{,} % to get "[numbered]" references from natbib
\smartqed  % flush right qed marks, e.g. at end of proof
\usepackage{graphicx}
\journalname{Few-Body Systems (EFB22)}
\newcommand{\eq}{\begin{eqnarray}}
\newcommand{\en}{\end{eqnarray}}

\begin{document}

\title{Electromagnetic $N-\Delta(1232)$ transitions
within the point-form of relativistic quantum mechanics
\thanks{This work is supported by National Sciences Foundations of China
Nos. 10975146, 11035006 and 11261130, as well as supported, in part,
by the DFG and the NSFC through funds provided to the Sino-Germen
CRC 110 ``Symmetries and the Emergence of Structure in QCD''.}}

%\subtitle{Do you have a subtitle?\\ If so, write it here}

%\titlerunning{Short form of title}        % if too long for running head

\author{Yubing Dong \and Mauro Giannini \and Elena Santopinto
\and Andrea Vassallo}

%\authorrunning{Short form of author list} % if too long for running head

\institute{Y. Dong\at
           Institute of High Energy Physics, Beijing 100049, People's Republic of China\\
           %\vspace*{.3\baselineskip}\\
           Theoretical Physics Center for Science Facilities (TPCSF), CAS,
           Beijing 100049, People's Republic of China\\
           \email{dongyb@ihep.ac.cn}\\           %  \\
%          \emph{Present address:} of F. Author  %  if needed
           \and Mauro Giannini \and Elena Santopinto \and Andrea Vassallo \at
           INFN, Sezione di Genova and universita di Genova, via Dodecaneso 33, I-16146, Genova,
           Italy}

\date{Received: date / Accepted: date}
% The correct dates will be entered by the editor

\maketitle

\begin{abstract}
The electromagnetic $N-\Delta(1232)$ transition amplitudes
are calculated using the point-form of relativistic quantum
mechanics. The relativistic effects incorporated in the
electromagnetic matrix elements give a good description of the
transition amplitudes to the $\Delta(1232)$
resonance, reproducing well the $Q^2$ behaviour of the data, apart
from the low $Q^2$ one.

\end{abstract}

\section{Introduction}

The study of nucleon electromagnetic form factors and the
electromagnetic transitions of the nucleon resonances is always of
great interest. It can give a detailed information on the internal
structure of the nucleon and its excitations. There has been a large
amount of model calculations in the past several decades, based on
both the non-relativistic and relativistic frameworks.  It is
expected that more accurate data to a higher $Q^2$ region will come
out with the 12 GeV upgraded JLab. facility. Therefore a more
precise description of the transition amplitudes in this region is
required.

In 1949, Dirac \cite{dirac} first proposed three equivalent forms of
the relativistic dynamics. They are the instant, light-front and point-forms.
Here, we use the point-form, since all the components of the four-momentum
$P_{\mu}$ ($\mu = 0, 1, 2, 3$) are associated with the interactions and
other operators, like the angular momentum and
Lorentz boost operators, are interaction free. Therefore, the advantage of
the point-form of relativistic quantum mechanics is that all the Lorentz
transformations remain purely kinematic and the theory is manifestly Lorentz
covariant.

People are more familiar with the instant and light-front forms than
the point-form, since the two were rather popular in the past
decades and most of the calculations were based on the two
frameworks. The point-form has been discussed by Keister and Polyzou
\cite{kp} and recently has been carefully and systematically studied
by Klink \cite{k}. It has also been employed in the calculations of
the nucleon form factors
\cite{grazpv1,grazpv2,grazpv3,melde,ge,ge2}, the resonance strong
decays \cite{graz-pd,graz-pd2}, and the form factors of pion and
deuteron \cite{dd,dd1,akp,akp1}. Those results show the importance
of the relativistic description of the systems, particularly when
the momentum transfer is large.

In this work, the point-form of relativistic quantum mechanics will
be employed to calculate the electromagnetic transition amplitudes
of the nucleon to $\Delta(1232)$. Here the wave functions of the
nucleon and its resonances from the hyper-central potential model
\cite{hyp} are employed. It is expected that the relativistic
description both for the wave functions and for the matrix elements
could well reproduce the $Q^2$-dependence of the transition
amplitudes. This work is organized as follows. In Sect. 2, the
relativistic hyper-central potential model will be briefly discussed
and the point-form of relativistic quantum mechanics is displayed
and applied to the study of the electromagnetic $N-\Delta(1232)$
transitions. Numerical results and a short summary will be given in
Sect 3.

\label{intro}

\section{Hyper-Central Potential Model and the Point-Form of Relativistic
Quantum Mechanics}

The hyper-central potential model was proposed a long time ago
\cite{hyp} and since then it has been used for the calculations of
the baryon electromagnetic properties
\cite{hyp-ff1,hyp-ff11,hyp-ff12,hyp-ff13,hyp-ff14}, in particular
for the predictions of the transition form factors of the nucleon to
its baryon resonances \cite{sg}.  The model has also been extended
to a relativistic version replacing the non-relativistic kinetic
operator by a fully relativistic one \cite{ge,ge2}.

The mass operator in the relativistic hypercentral constituent quark
model is given by \cite{ge,ge2} \eq {\hat M}=\sum_{i=1}^3
\sqrt{m^2+\vec{k}_i^{~2}}-\frac{\tau}{x}+\alpha x +M_{hyp}. \en In
our calculation, the center-of-mass frame is considered and thus
$\sum^3_{i=1}\vec{k}_i=0$. In Eq. (1), the hyper-radius
$x=\sqrt{\vec{\rho}^{~2}+\vec{\lambda}^{~2}}$ with
$\vec{\rho}=\frac{1}{\sqrt{2}}(\vec{r}_1-\vec{r}_2)$ and
$\vec{\lambda}=\frac{1}{\sqrt{6}}(\vec{r}_1+\vec{r}_2-2\vec{r}_3)$
being the internal Jacobi coordinates. $M_{hyp}$ is the hyperfine
interaction, which is spin-dependent. The spin-independent part of
the interaction includes, at least in some sense, the three-body
interactions. It is different from the other ordinary constituent
quark models where only the two-body interactions are taken into
account. On the other hand, it may be considered as the hypercentral
approximation to the two-body potential. The relativistic mass
operator can be diagonalized by means of a variational method and
one has to work in the momentum space due to the relativistic
kinetic energy operator.

In the point-form of relativistic quantum mechanics, in order to
construct the interacting four-momentum operator, one usually uses
the Bakamjian-Thomas method \cite{BT} by putting the interactions
into the mass operator $\hat{{\cal M}}$. Thus, $\hat{{\cal M}}$ is
divided into two parts. One is the interaction free mass operator
$\hat{{\cal M}}_{fr}$ and another is the interacting mass operator
$\hat{{\cal M}}_{int}$. The four-momentum $P^{\mu}$ is related to
the mass operator by \eq P^{\mu}=\hat{{\cal M}}V^{\mu}_{fr}, \en
where the four-velocity operator $V^{\mu}_{fr}$ is interaction free.
According to the commutation relations satisfied by the operators of
the dynamical system and to the fact that $P^{\mu}$ is a Lorentz
vector, one gets the relation of $[V^{\mu}_{fr},\hat{{\cal M}}] = 0$
and $\hat{{\cal M}}$ is a Lorentz scalar. Therefore, the eigenstates
of the four-momentum operator are the eigenstates of both the mass
and the velocity operators. In the center-of-mass frame, we can
obtain the wave functions of the three-quark system by solving a
relativistic Schr\"{o}dinger equation. The obtained wave functions
are the eigenstates of the mass operator with interactions.

In the point-form of relativistic quantum mechanics, the Lorentz
transformations remain purely kinematic, namely, they are
interaction free. The so-called velocity state is usually introduced
as follows \cite{k}, \eq \mid
v;\vec{k}_1,\vec{k}_2,\vec{k}_3;\mu_1,\mu_2,\mu_3>
&=&U_{B(v)}\mid\vec{k}_1,\vec{k}_2,\vec{k}_3;\mu_1,\mu_2,\mu_3>\\
\nonumber &=&\Pi_{i=1}^3D^{1/2}_{\sigma_i\mu_i}\big
[R_W(k_i,B(v)\big] \mid
\vec{p}_1,\vec{p}_2,\vec{p}_3;\sigma_1,\sigma_2,\sigma_3> \en where
$k_i$ (with $i =1,2,3$) are the quark momenta in the center-of-mass
system, $B(v)$ is a Lorentz boost with four-velocity $v$,
$p_i=B(v)k_i$, and $U_{B(v)}$ is a unitary representation of $B(v)$.
$D^{1/2}(RW)$ is the spin-1/2 representation matrix of the Wigner
rotation. It has been proved \cite{k} that all the Wigner rotations
of a canonical boost of a velocity state are the same, and thus the
spins can be coupled together to the total spin of the state as in
the non-relativistic framework as well as in the center-of-mass
frame. This is the practical advantage of using the velocity state.

To calculate the photo- and electro-production amplitudes of the
nucleon resonances, we simply employ the point-form spectator
impulse approximation for the electromagnetic interaction. The
current operator is assumed to be the single-particle one
\cite{grazpv1,grazpv2,grazpv3,melde}, \eq <p_i',\lambda_i'\mid
j^{\mu}\mid p_i,\lambda_i>=e_i\bar{u}(p_i',\lambda_i')
\gamma^{\mu}u(p_i,\lambda_i), \en where $u(p_i,\lambda_i)$ is the
Dirac spinor with momentum $p_i$ and spin $\lambda_i$ for the $i$-th
struck quark. \label{sec:2}

\section{Numerical results and summary}

In this work, we present the preliminary results for the
electromagnetic transition amplitudes of the $N-\Delta(1232)$ based
on the point-form of relativistic quantum mechanics. Here we employ
the wave functions of the nucleon and the nucleon resonance
$\Delta(1232)$ obtained from the relativistic hyper-central
potential model. Figure 1 reports the obtained transverse transition
amplitudes to the $\Delta(1232)$. In the figure the data, from Refs.
\cite{azn, MAID}, are also shown for a comparison. We can see that
our present framework can well reproduce the transverse transition
amplitudes of the $\Delta(1232)$ resonance in the region of
$Q^2>~1~GeV^2$. The present calculation with the point-form of
relativistic quantum mechanics is an improvement with respect to the
calculations with the non-relativistic hyper-central potential
model. For the amplitudes in the small $Q^2$ region, it is expected
that the quark-antiquark pair production mechanism plays a dominant
role. In order to take into account this effect in a consistent way
one has to work within the unquenched quark model
\cite{uqm,uqm1,uqm2}.

\begin{figure}[t]
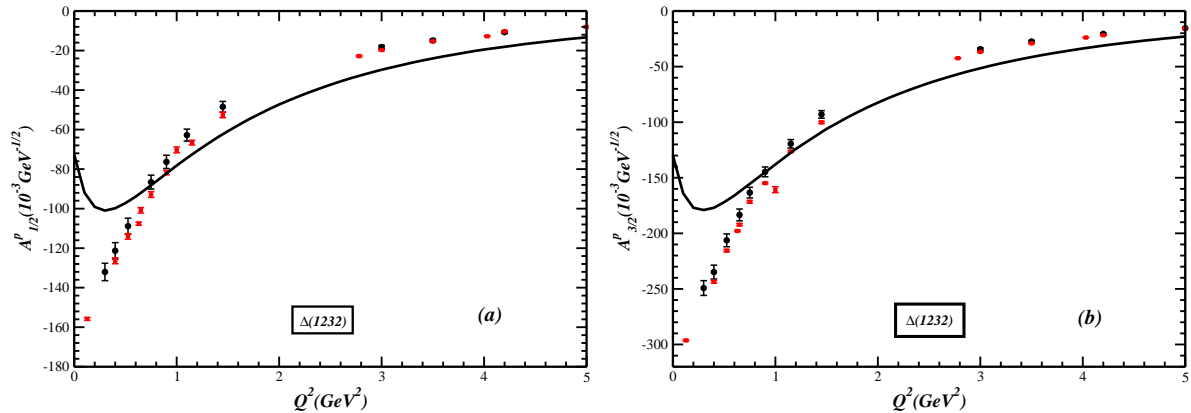

\vspace{0.5cm}
\begin{minipage}[b]{0.5\linewidth}
%\centering \includegraphics[width=0.95\textwidth]{gda12qge.eps}
\centering \includegraphics[width=0.98\textwidth]{fig1.eps}
\end{minipage}
\begin{minipage}[b]{0.5\linewidth}
%\centering \includegraphics[width=0.95\textwidth]{gda32qge.eps}
\centering \includegraphics[width=0.98\textwidth]{fig2.eps}
\end{minipage}
\caption{The preliminarly estimated $N-\Delta(1232)$ transverse
transition amplitudes (solid curves), (a) for $A_{1/2}$ and (b) for
$A_{3/2}$. The date are from Refs. \cite{azn,MAID}.}
%\end{minipage}
\end{figure}

To summarize, we have applied the wave functions of the nucleon and
$\Delta(1232)$, obtained from the relativistic hyper-central
potential model, for the calculation of the electromagnetic
$N-\Delta(1232)$ transition amplitudes based on the point-form of
relativistic quantum mechanics. Our numerical results show
explicitly the advantage of the present fully relativistic
description in the region of $Q^2 > 1 GeV^2$. Other electromagnetic
observables of the nucleon resonances, like the transition
amplitudes to the low-lying nucleon resonances of $S_{11}(1535)$ and
$D_{13}(1520)$, will be published elsewhere.

\begin{acknowledgements}
One of Author (YBD) thanks the Department of Physics, Genova University,
for the warm hospitality.
\end{acknowledgements}


\begin{thebibliography}{99}

\bibitem{dirac}
Dirac, P. A. M.: Forms of Relativistic Dynamics. Rev. Mod. Phys.
{\bf 21}, 392-399 (1949)

\bibitem{kp}
Keister, B. D, Polyzou, W.N.: In: Negele, J. W, Vogt, E. W. (eds.)
Advanced Nuclear Physics, vol. {\bf 21} pp.225-479. Plenum (1991)

\bibitem{k}
Klink, W. H.: Point form relativistic quantum mechanics and
electromagnetic form factors. Phys. Rev. C. {\bf 58} 3587-3604,
(1998)

\bibitem{grazpv1}
Wagenbrunn, R. F., Boffi, S., Klink, W., Plessas, W., Radici, R.:
Covariant nucleon electromagnetic form-factors from the Goldstone
boson exchange quark model. Phys. Lett. B. {\bf 511}, 33-39 (2001)

\bibitem{grazpv2}Glozman, L. Y., Radici, M., Wagenbrunn, R. F., Boffi, S.,
Klink, W., Plessas, W.: Covariant axial form-factor of the nucleon
in a chiral constituent quark model. Phys. Lett. B. {\bf 516},
183-190 (2001)

\bibitem{grazpv3}
Boffi, S., Glozman, L. Y., Klink, W., Plessas, W., Radici, M.,
Wagenbrunn, R. F.: Covariant electroweak nucleon form-factors in a
chiral constituent quark model. Eur. Phys. J. A. {\bf 14}, 17-21
(2002)

\bibitem{melde} Melde, T., Berger, K., Canton, L., Plessas, W., Wagenbrum,
R. F.: Electromagnetic nucleon form factors in instant and point
form. Phys. Rev. D. {\bf 76}, 074020 (2007)

\bibitem{ge}
De Sanctis, M., Giannini, M. M., Santopinto, E., Vassallo, A.:
Electromagnetic form factors and the hypercentral constituent quark
model. Phys. Rev. C. {\bf 76}, 062201 (2007)

\bibitem{ge2}Santopinto, E., Vassallo, A., Giannini, M. M., De Sanctis, M.:
High $Q^2$ behavior of the electromagnetic form factors in the
relativistic hypercentral constituent quark model. Phys. Rev. C.
{\bf 82} 065204 (2010)

\bibitem{graz-pd}
Melde, T., Plessas, W., Wagenbrunn, R. F.: Covariant calculation of
mesonic baryon decays. Phys. Rev. C. {\bf 72}, 015207 (2005)

\bibitem{graz-pd2}Melde. T., Canton. L., Plessas. W.: Structure of meson-baryon
interaction vertices. Phys. Rev. Lett. {\bf 102},  132002 (2008)

\bibitem{dd}
Amghar, A., Desplanques, B., Theussl, L.: The form-factor of the
pion in point form of relativistic dynamics revisited. Phys. Lett.
B. {\bf 574}, 201-209 (2003)

\bibitem{dd1}Desplanques, B., Dong, Y. B.: RQM description of PS meson
form factors, Constraints from space-time translations, and
underlying dynamics. Eur. Phys. J. A. {\bf 47}, 13-33 (2011)

\bibitem{akp}
Allen, T. W., Klink, W. H.: Pion charge form-factor in point form
relativistic dynamics. Phys. Rev. C. {\bf 58}, 3670-3673 (1998)

\bibitem{akp1}
Allen, T. W., Klink, W. H., Polyzou, W. N.: Point form analysis of
elastic deuteron form factors. Phys. Rev. C. {\bf 63}, 034002 (2011)

\bibitem{hyp}
Ferraris, M., Giannini, M. M., Pizzo, M., Santopinto, E., Tiator,
L.: A three body force model for the baryon. Phys. Lett. B. {\bf
364}, 231-238 (1995)

\bibitem{hyp-ff1}
Aiello, M., Ferraris, M., Giannini, M. M., Pizzo, M., Santopinto,
E.: A three body force model for the electromagnetic excitation of
the nucleon. Phys. Lett. B. {\bf 387}, 215-221 (1996)

\bibitem{hyp-ff11}Aiello, M., Giannini, M. M., Santopinto, E.: Electromagnetic
transition form-factors of negative parity nucleon resonances. J.
Phys. G. {\bf 24}, 753-762 (1998)

\bibitem{hyp-ff12}
De Sanctis, M., Santopinto, E., Giannini, M. M.: A relativistic
study of the nucleon form-factors. Eur. Phys. J. A. {\bf 1}, 187-192
(1998)

\bibitem{hyp-ff13}De Sanctis, M., Santopinto, E., Giannini, M. M.:
A relativistic study of the nucleon helicity amplitudes. Eur. Phys.
J. A. {\bf 2}, 403-409 (1998)

\bibitem{hyp-ff14}De Sanctis, M., Giannini, M. M., Repetto, L., Santopinto, E.:
Proton form-factors in the hypercentral constituent quark model.
Phys. Rev. C. {\bf 62}, 025208 (2000)

\bibitem{sg}
Santopinto, E., Giannini, M. M.: Systematic study of longitudinal
and transverse helicity amplitudes in the hypercentral constituent
quark model. Phys. Rev. C. {\bf 86}, 065202 (2012)

\bibitem{BT}
Bakamjian, B., Thomas, L. H.: Relativistic particle dynamics 2.
Phys. Rev. {\bf 92}, 1300-1310 (1953)

\bibitem{azn}
Aznauryan, I. G., et al.: (CLAS Collaboration) Electro-excitation of
nucleon resonances from CLAS data in single pion electro-production.
Phys. Rev. C. {\bf 80}, 055203 (2009)

\bibitem{MAID}
Drechsel, D., Kamalov, S. S., Tiator, L.: Unitary isobar model-MAID
2007. Eur. Phys. J. A. {\bf 34}, 69-97 (2007)

\bibitem{uqm}
Bijker, R., Santopinto, E.: Unquenched quark model for baryons:
Magnetic moments, spins and orbital angular momenta. Phys.Rev. C.
{\bf 80},  065210 (2010)

\bibitem{uqm1}Santopinto, E., Bijker, R.: Flavor asymmetry of sea
quarks in the unquenched quark model Phys. Rev. C. {\bf 82}, 062202
(2010)

\bibitem{uqm2}Bijker, R., Ferretti, J., Santopinto, E.: s\={s} sea pair
contribution to electromagnetic observables of the proton in the
unquenched quark model Phys.Rev. C. {\bf 85}, 035204 (2012)


\end{thebibliography}
\end{document}